
\input phyzzx
\hoffset=1truein
\voffset=1.0truein
\hsize=6truein
\def\TITLEPAGE{\frontpagetrue}
\def\CALT#1{\hbox to\hsize{\tenpoint \baselineskip=12pt
        \hfil\vtop{\hbox{\strut CALT-68-#1}
        \hbox{\strut DOE RESEARCH AND}
        \hbox{\strut DEVELOPMENT REPORT}}}}

\def\CALTECH{\smallskip
        \address{California Institute of Technology, Pasadena, CA 91125}}

\def\AUTHOR#1{\vskip .5in \centerline{#1}}

\def\ABSTRACT#1{\vskip .5in \vfil \centerline{\twelvepoint \bf Abstract}
        #1 \vfil}
\def\ENDTITLEPAGE{\vfil\eject\pageno=1}

\def\sqr#1#2{{\vcenter{\hrule height.#2pt
      \hbox{\vrule width.#2pt height#1pt \kern#1pt
        \vrule width.#2pt}
      \hrule height.#2pt}}}

\def\section#1#2{
\noindent\hbox{\hbox{\bf #1}\hskip 10pt\vtop{\hsize=5in
\baselineskip=12pt \noindent \bf #2 \hfil}\hfil}
\medskip}

\def\underwig#1{        
        \setbox0=\hbox{\rm \strut}
        \hbox to 0pt{$#1$\hss} \lower \ht0 \hbox{\rm \char'176}}

\def\bunderwig#1{       
        \setbox0=\hbox{\rm \strut}
        \hbox to 1.5pt{$#1$\hss} \lower 12.8pt
         \hbox{\seventeenrm \char'176}\hbox to 2pt{\hfil}}

\def\MEMO#1#2#3#4#5{
\frontpagetrue
\centerline{\tencp INTEROFFICE MEMORANDUM}
\smallskip
\centerline{\bf CALIFORNIA INSTITUTE OF TECHNOLOGY}
\bigskip
\vtop{\tenpoint
\hbox to\hsize{\strut \hbox to .75in{\caps to:\hfil}\hbox to 3.8in{#1\hfil}
\quad\the\date\hfil}
\hbox to\hsize{\strut \hbox to.75in{\caps from:\hfil}\hbox to 3.5in{#2\hfil}
\hbox{{\caps ext-}#3\qquad{\caps m.c.\quad}#4}\hfil}
\hbox{\hbox to.75in{\caps subject:\hfil}\vtop{\parindent=0pt
\hsize=3.5in #5\hfil}}
\hbox{\strut\hfil}}}

\tolerance=10000
\hfuzz=5pt

\TITLEPAGE
\CALT{1798}
\setbox0=\hbox{\strut DOE RESEARCH AND}
\hbox to \hsize{\hfil\vtop{\hbox{UCSD/PTH 92-20}
\hbox to\wd0{\hfil}}}
\bigskip
\titlestyle {Two Photon Contribution to Polarization in $K^+ \rightarrow
\pi^+ \mu^+ \mu^-$~~~~~~\foot{Work supported in part by the U.S. Dept. of
Energy
under Contracts DEAC-03-81ER40050 and DE-FG03-90ER40546 and NSF grant
PHY-9057135.}}
\AUTHOR{Ming Lu and Mark B. Wise}
\CALTECH
\medskip
\AUTHOR{Martin J. Savage}
\medskip
\centerline{{\it Department of Physics, University of California, San
Diego,}}
\centerline{{\it 9500 Gilman Drive, La Jolla, CA ~92093-0319}}
\ABSTRACT{Short distance
physics involving virtual top and charm quarks contributes to $\mu^+$
(and $\mu^-$) polarization in the decay $K^+ \rightarrow \pi^+ \mu^+
\mu^-$.
Measurement of the parity violating
asymmetry $(\Gamma_R - \Gamma_L)/(\Gamma_R + \Gamma_L)$, where
$\Gamma_R$ and $\Gamma_L$ are the rates to produce right and left-handed
$\mu^+$, may provide valuable information on the unitarity triangle.
The parity violating
asymmetry also gets a contribution from Feynman diagrams with two
photon intermediate states.  We estimate this two photon contribution to
the asymmetry and
discuss briefly the two photon contribution to time reversal odd
asymmetries that involve both the $\mu^+$ and $\mu^-$ polarizations.}

\ENDTITLEPAGE

\eject

\noindent {\bf 1.  Introduction}

In the minimal standard model the coupling of the quarks to the
$W$-bosons has the form
$$      {\cal L}_{int} = - {g_2\over \sqrt{2}} \bar u_L^j \gamma_\mu
V^{jk} d_L^k W^\mu + h.c. \,\, . \eqno (1)$$
Here the repeated generation indices $j$ and $k$ are summed over 1,2,3
and $g_2$ is the weak SU(2) gauge coupling.  $V$ is a $3\times 3$
unitary matrix that arises from diagonalization of the quark mass
matrices.  By redefining the phases of the quark fields it is possible
to write $V$ in terms of the four angles $\theta_1, \theta_2, \theta_3$
and $\delta$ (For $N_g$ generations there are $(N_g - 1)^2$ angles.)  The
$\theta_j$ are analogous to the Euler angles and $\delta$ is a phase
that gives rise to CP violation.  Explicitly$^{[1]}$
$$      V = \pmatrix{c_1 & -s_1 c_3 & -s_1 s_3\cr
s_1 c_2 & c_1 c_2 c_3 - s_2 s_3 e^{i\delta} & c_1 c_2 s_3 + s_2 c_3
e^{i\delta}\cr
s_1 s_2 & c_1 s_2 c_3 + c_2 s_3 e^{i\delta} & c_1 s_2 s_3 - c_2 c_3
e^{i\delta}\cr}\eqno (2)$$
where $c_i \equiv \cos \theta_i$ and $s_i \equiv \sin \theta_i$.  It is
possible to choose the $\theta_j$ to lie in the first quadrant by
redefining the quark fields.  Then the quadrant of $\delta$ has physical
significance and cannot be chosen by convention.  A value of $\delta$
not equal to zero or $\pi$ gives rise to CP violation.

Experimental information on nuclear $\beta$ decay and weak decays of
kaons, hyperons and $B$ mesons shows that the angles
$\theta_j$ are small (but different from zero).  The angle
$\theta_1$ is essentially the Cabibbo angle.  It is by far the best
known of the angles$^{[2]}$
$$      \sin \theta_1 = 0.22 \,\, , \eqno (3)$$
(with an error at the percent level).

Unitarity of the Cabibbo-Kobayashi-Maskawa matrix $V$ gives that
$$      V_{ud} V^*_{ub} + V_{cd} V^*_{cb} + V_{td} V^*_{tb} = 0 \,\, .
\eqno (4)$$
We can think of each of the three complex numbers $(V_{ud} V^*_{ub}$,
etc.) on the l.h.s. of eq. (4) as vectors in the complex plane.  These
vectors add to zero and so by translating them they
form the sides of a triangle that is often called the unitarity
triangle.  With the parametrization of the Cabibbo-Kobayashi-Maskawa
matrix in eq. (2) we have
$$      \eqalignno{V_{ud} V_{ub}^* &\simeq - s_1 s_3 & (5a)\cr
        V_{td} V_{tb}^* &\simeq - s_1 s_2 e^{-i\delta} & (5b)\cr
        V_{cd} V_{cb}^* &\simeq s_1 (s_3 + s_2 e^{-i\delta}) & (5c)\cr}$$
The unitarity triangle specifies the angles $\theta_2, \theta_3$ and
$\delta$.  From eqs. (5) it is clear that the length of two sides gives
$\theta_2$ and $\theta_3$ while the angle between two of the sides is
$\pi - \delta$.

The orientation of the unitarity triangle in the complex plane depends
on the phase convention in the Cabibbo-Kobayashi-Maskawa matrix.  The length of
the sides and the angles at each vertex $\alpha, \beta, \gamma$ are
independent of the phase convention.  When there is no CP violation the
unitarity triangle collapses to a line.  One common orientation for the
triangle has $V_{cd} V_{cb}^*$ lying along the real axis.  It is
conventional to rescale the side on the real axis to unit length and
locate one vertex at the origin of the complex plane.  This is shown in
Fig. 1.  With this convention the unitarity triangle is specified by the
coordinates in the complex plane, $\rho + i\eta$, of the vertex
associated with the angle $\alpha$.

It is important to determine the unitary triangle by measuring
quantities that do not violate CP.  The resulting values of the
weak mixing angles can then be used to predict the expected values of CP
violating quantities.  In this way the standard six-quark model for CP
violation can be tested.  At the present time it is not known if the CP
violation observed in kaon decays is due to the phase in the
Cabibbo-Kobayashi-Maskawa matrix or from new physics, beyond that in the
minimal standard model, or both.

$B$-meson decays give valuable information on the unitarity triangle.
However, rare kaon decays where a virtual top quark plays an important
role can also be useful.  For example, an accurate measurement of the
branching ratio for $K^+ \rightarrow \pi^+ \nu \bar\nu$ would restrict
the $\alpha$ corner of the unitarity triangle to lie on a circle.

In Ref. [3] it was pointed out that the measurement of polarization in
$K^+ \rightarrow \pi^+ \mu^+ \mu^-$ decay can also lead to valuable
information on the weak mixing angles.  The dominant contribution to the
$K^+ \rightarrow \pi^+ \mu^+\mu^- $ decay amplitude comes from Feynman
diagrams where a single photon produces the $\mu^+ \mu^-$ pair.  Even
though the weak interactions violate parity maximally  the one photon part
of the decay amplitude is necessarily parity conserving and doesn't
contribute to the parity violating asymmetry $\Delta_{LR} =
(\Gamma_R - \Gamma_L)/(\Gamma_R + \Gamma_L)$, where $\Gamma_R$ and
$\Gamma_L$ are the rates to produce right and left-handed $\mu^+$
respectively.\foot{By right- (or left-) handed we mean that the spin is
directed along (or opposite) the direction of motion, i.e., helicity $+
1/2$  (or $- 1/2$).}  This parity violating asymmetry arises
predominantly from two sources:

\item{(i)}  the interference of $W$-box and $Z$-penguin Feynman diagrams
(see Figure 2) with the one-photon piece.

\item{(ii)}  the interference of Feynman diagrams where two photons
create the $\mu^+ \mu^-$ pair with the one photon piece.

\noindent If the short distance $W$-box and $Z$-penguin part dominates
the asymmetry then its measurement can lead to important information on
the unitarity triangle.  The main purpose of this paper is to examine the
long-distance two photon contribution to the $K^+ \rightarrow \pi^+ \mu^+
\mu^-$ decay amplitude and in particular its influence on the parity
violating asymmetry $\Delta_{LR}$.
Ref. [3] also noted that there are $T$-odd asymmetries which involve
both the $\mu^+$ and $\mu^-$ polarizations and can arise from the
interference of the $Z$-penguin and $W$-box Feynman diagrams with the
one photon piece.  Detailed predictions for the short distance
contribution to these $T$-odd asymmetries were made in Ref. [4].  Here we
stress that the $T$-odd asymmetries also receive a contribution from the
interference of the absorptive part of the parity violating two photon
contribution with the one photon piece.

\noindent {\bf 2.  Kinematics}

The dominant part of the $K^+ \rightarrow \pi^+ \mu^+ \mu^-$ decay
amplitude comes from Feynman diagrams where a single photon produces the
$\mu^+\mu^-$ pair.  The one photon contribution to the invariant matrix
element has the form
$$      {\cal M}^{(pc)} = {s_1 G_F\over\sqrt{2}} \alpha f(s) (p_K + p_\pi)^\mu
\bar u (p_-, s_-) \gamma_\mu v(p_+ , s_+) \,\, , \eqno (6)$$
where $p_K$ and $p_\pi$ are the four-momentum of the kaon and pion and $p_\pm$
are the
four-momenta of the $\mu^\pm$.  In eq. (6) $s_\pm$ are the spin vectors
for the $\mu^\pm$ while $\sqrt{s}$ is the invariant mass of the
$\mu^+\mu^-$ pair
$$      s = (p_+ + p_-)^2 \,\, . \eqno (7)$$
We shall parametrize the differential decay rate in terms of $s$ and
$\theta$ the angle between the three-momentum of the kaon and the
three-momentum of the $\mu^-$ in the $\mu^+\mu^-$ pair rest frame.  In
terms of these variables the inner products of four-momenta are
$$      p_- \cdot p_+ = s/2 - m_\mu^2 \,\, , \eqno (8a)$$
$$      (p_K + p_\pi)^2 = 2(m_K^2 + m_\pi^2) - s \,\, ,\eqno (8b)$$
$$      2p_+ \cdot (p_K + p_\pi) = (m_K^2 - m_\pi^2) +  \sqrt{1 -
{4 m_\mu^2\over s}} [ (s + m_K^2 - m_\pi^2)^2 - 4s
m_K^2)]^{1/2} \cos\theta \,\, . \eqno (8c)$$
For a right or left-handed $\mu^+$ the dot products of the polarization
four-vector $s_+^\mu$ with the $\mu^-$ and kaon four-momenta are
$$      s_+^{(R)} \cdot p_- = - s_+^{(L)} \cdot p_- = {s\over 2m_\mu}
\sqrt{1 - {4m_\mu^2\over s}} \,\, , \eqno (9a)$$
$$      s_+^{(R)} \cdot p_K = - s_+^{(L)} \cdot p_K = {1\over 4m_\mu}
\bigg\{ \sqrt{1- {4m_\mu^2\over s}} (s + m_K^2 - m_\pi^2)$$
$$      + [(s + m_K^2 - m_\pi^2)^2 - 4sm_K^2]^{1/2} \cos\theta \bigg\}
\,\, . \eqno (9b)$$

The total differential decay rate is dominated by the one photon piece
and the invariant amplitude in eq. (6) gives
$$      d(\Gamma_R + \Gamma_L)/d\cos\theta ds = {s_1^2 G_F^2 \alpha^2
|f(s)|^2\over 2^9 m_K^3 \pi^3} \sqrt{1- {4m_\mu^2\over s}} [(m_K^2 -
m_\pi^2 + s)^2 - 4sm_K^2]^{3/2}$$
$$      \left [ 1 - \left(1 - {4m_\mu^2\over s}\right)
\cos^2\theta\right]\,\, . \eqno (10)$$

The parity violating part of the decay amplitude has the form
$$      {\cal M}^{(pv)} = {s_1 G_F \alpha\over\sqrt{2}} [B(p_K + p_\pi)^\mu + C
(p_K - p_\pi)^\mu]$$
$$      \cdot \bar u (p_-, s_-) \gamma_\mu \gamma_5 v (p_+, s_+) \,\, .
\eqno (11)$$
The parameters $B$ and $C$ in eq. (11) get contributions from the
$Z$-penguin and $W$-box Feynman diagrams as well as from Feynman
diagrams with two photons.

The difference in decay amplitudes for right and left-handed $\mu^+$
arises from the interference of the parity conserving part of the decay
amplitude in eq. (6) with the parity violating part of the decay
amplitude in eq. (11).  This gives
$$      d(\Gamma_R - \Gamma_L)/d\cos \theta ds = {-s_1^2 G_F^2
\alpha^2\over 2^8 m_K^3 \pi^3} \sqrt{1 - {4m_\mu^2\over s}} [(s + m_K^2
- m_\pi^2)^2 - 4sm_K^2]$$
$$      \Bigg\{ Re (f^*(s) B) \sqrt{1- {4m_\mu^2\over s}} [(s + m_K^2 -
m_\pi^2) - 4sm_K^2]^{1/2} \sin^2 \theta$$
$$      + 4 \left[Re (f^*(s) B)\left({m_K^2 - m_\pi^2\over s}\right) +
Re (f^*(s)C)\right] m_\mu^2 \cos \theta \Bigg\} \,\, . \eqno (12)$$
Note that in eq. (12) the contribution of $C$ vanishes when the
difference of decay rates is integrated over $\theta$.

\noindent {\bf 3.  The Parity Conserving Amplitude}

The parity conserving amplitude arises predominantly from Feynman
diagrams where a single photon produces the $\mu^+ \mu^-$ pair.  It is
characterized by the function $f(s)$ introduced in eq. (6) of Section
2.  The absolute value of $f(s)$ has been determined by experimental
data on $K^+ \rightarrow \pi^+ e^+ e^-$.  A good fit to the differential
decay rate is obtained from$^{[5]}$
$$      |f(s)| = |f(0)| (1 + \lambda s/m_\pi^2) \,\, , \eqno (13)$$
with $\lambda = 0.11$ and $|f(0)| = 0.31$.

Using chiral perturbation theory, the imaginary part of $f(s)$, arises
from the Feynman diagrams in Fig. 3, with the pions in the loop on their
mass shell.  The strong interactions of the pseudo-Goldstone bosons $\pi,
K$ and $\eta$ are described by the effective chiral Lagrangian
$$      {\cal L} = {f^2\over 8} Tr \partial_\mu \Sigma \partial^\mu
\Sigma^{\dagger} + v Tr (m_q \Sigma + \Sigma^{\dagger} m_q) + ... \,\, .
\eqno (14)$$
In eq. (14) $f$ is the pion decay constant, $f\simeq 132$MeV and $m_q$
is the quark mass matrix
$$      m_q = \left[\matrix{m_u & 0 & 0\cr
0 & m_d & 0 \cr
0 & 0 & m_s\cr}\right]\eqno (15)$$
The pseudo-Goldstone boson fields occur in the $3\times 3$ special
unitary matrix $\Sigma$.  Explicitly
$$      \Sigma = \exp (2iM/f) \,\, , \eqno (16)$$
where
$$      M = \left[\matrix{\pi^0/\sqrt{2} + \eta/\sqrt{6} & \pi^+ & K^+\cr
\pi^- & -\pi^0/\sqrt{2} + \eta/\sqrt{6} & K^0\cr
K^- & \bar K^0 & -2\eta/\sqrt{6}\cr}\right] \,\, . \eqno (17)$$
In Fig. 3 a shaded circle denotes an interaction vertex arising from the
strong interaction effective Lagrangian density in eq. (14).  The
effective Lagrangian for $\Delta s = 1$ weak nonleptonic decays transforms
under chiral $SU(3)_L \times SU(3)_R$ as $(8_L, 1_R) + (27_L, 1_R)$.  In
terms of $\Sigma$ the $(8_L, 1_R)$ part of the effective Lagrangian
density\foot{It dominates over the $(27_L, 1_R)$ part of the Lagrangian.}
for weak nonleptonic kaon decays is
$$      {\cal L} = g_8 {G_F\over 4\sqrt{2}} s_1 f^4 Tr O_W
\partial^\mu \Sigma \partial_\mu \Sigma^{\dagger} + ... \,\, , \eqno
(18)$$
where
$$      O_W = \left[\matrix{0 & 0 & 0\cr
0 & 0 & 0\cr
0 & 1 & 0\cr}\right] \,\, . \eqno (19)$$
The measured $K_S \rightarrow \pi^+ \pi^-$ decay rate implies
that$^{[6]}$
$|g_8| \simeq 5.1$.  In Fig. 3 a shaded square denotes an interaction
vertex from the effective Lagrangian in eq. (18).  The Feynman diagrams
in Fig. 3 give$^{[6]}$
$$      Im f(s) = - (g_8/24) (1 - 4 m_\pi^2/s)^{3/2} \theta (s - 4
m_\pi^2) \,\, . \eqno (20)$$
The imaginary part of $f(s)$ is largest at the maximum value of $s,
s_{max} = (m_K - m_\pi)^2$.  Eq. (20) implies that (up to a sign)
$Imf(s_{max}) \simeq
0.05$ and so the imaginary part of $f(s)$ is expected to be much smaller
than its real part.

Chiral perturbation theory also predicts $Ref(s)$ up to a $s$ independent
constant that is determined by the total decay rate.$^{[6]}$  The measured $s$
dependence given in eq. (13) is somewhat greater than what chiral perturbation
theory
gives but the experimental error is still quite large, i.e., $\lambda = 0.105
\pm 0.035 \pm 0.015$.

\noindent {\bf 4.  Short Distance Contribution to the Parity
Violating Amplitude}

The $Z$-penguin and $W$-box Feynman diagrams contribute to both $B$ and
$C$ of the parity violating amplitude in eq. (11).  Explicitly
$$      B = f_+ (s)\xi \qquad \qquad C = f_- (s)\xi\,\, , \eqno (21)$$
where $f_+ (s)$ and $f_-(s)$ are the form factors for $K_{\ell 3}$
semileptonic decay.  Conventionally their $s$-dependence is parametrized
by  $f_\pm (s) = f_\pm (0) (1 + \lambda_\pm s/m_\pi^2)$.
We use$^{[2]}$ $f_+ (0) = 1.02, \lambda_+ = 0.03, f_- (0) = -0.17$ and
$\lambda_- =  0$.
$\xi$ is a quantity that, apart from mixing angles, is essentially the same
as occurs in $B\rightarrow X_s e^+e^-$.  As noted in Ref. [3] it is given by
$$      \xi \simeq - \tilde{\xi}_c + \left({V_{ts}^* V_{td}\over
V_{us}^* V_{ud}}\right) \tilde{\xi}_t \,\, , \eqno (22)$$
where
$$      \tilde{\xi}_q = \tilde{\xi}_q^{(Z)} + \tilde{\xi}_q^{(W)}\,\, ,
\eqno (23)$$
is the sum of the contributions of the $Z$-penguin (superscript $Z$) and
$W$-box (superscript $W$).  In eq. (23)
$$      \tilde{\xi}_t^{(Z)} = {x\over \sin^2 \theta_W} ~ {1\over 16\pi}
\left[{(x - 6) (x - 1) + (3 x + 2) \ln x\over (x - 1)^2}\right] \eqno
(24a)$$
$$      \tilde{\xi}_t^{(W)} = {x\over\sin^2 \theta_W} ~ {1\over 8\pi}
\left[{x - 1 - \ln x\over (x - 1)^2}\right] \,\, , \eqno (24b)$$
with $x = m_t^2/M_W^2$ and
$$      \tilde{\xi}_c^{(Z)} \simeq {\eta^{(Z)}\over \sin^2 \theta_W} ~
{1\over 8\pi} \left({m_c^2\over M_W^2}\right) \ln (m_c^2/M_W^2) \eqno
(25a)$$
$$      \tilde{\xi}_c^{(W)} \simeq - {\eta^{(W)}\over \sin^2 \theta_W} ~
{1\over 8\pi} \left({m_c^2\over M_W^2}\right) \ln (m_c^2/M_W^2) \,\, .
\eqno (25b)$$
The QCD correction factors $\eta^{(Z)}$ and $\eta^{(W)}$ have been
computed in the leading logarithmic approximation$^{[7]}$ and they have the
values $\eta^{(Z)} \simeq 0.3$ and $\eta^{(W)} \simeq 0.6$.  Using $m_c =
1.5$ GeV and $M_W = 81$ GeV and $\sin^2 \theta_W = 0.23$  equations
(25a) and (25b) imply that $\tilde \xi_c = 1.4 \times 10^{-4}$.  The
value of $\tilde{\xi}_t$ depends sensitively on the top quark mass.
For $m_t = 140$ GeV, $\tilde{\xi}_t \simeq 0.51$ and for $m_t = 200$
GeV, $\tilde{\xi}_t \simeq 0.89$.

The coefficient of $\tilde{\xi}_t$ depends on the weak mixing angles.  It
is convenient to reexpress this combination of elements of the
Cabibbo-Kobayashi-Maskawa matrix in terms of $|V_{cb}|$ and the complex
coordinates $\rho + i \eta$ of the $\alpha$ vertex of the unitarity
triangle,
$$      V_{ts}^* V_{td}/V_{us}^* V_{ud} = (\rho - 1 + i\eta) | V_{cb}|^2
\,\, . \eqno (26)$$

The value of $|V_{cb}|$ can be obtained from the semileptonic decays $B
\rightarrow D^* e \bar\nu_e$ and $B \rightarrow D e\bar\nu_e$.  Using
heavy quark spin-flavor symmetry the hadronic form factors for these
decays can be expressed in terms of a single universal function of
``velocity-transfer.''$^{[8]}$  Furthermore, the normalization  of this
universal function is fixed at zero  recoil where both the $B$ and $D^*$
or $D$ are at rest.$^{[8,9,10]}$  Comparison of the predictions of heavy quark
symmetry with experimental data on these decays gives,$^{[11]}$ $|V_{cb}| =
0.043
\pm 0.007$.  At zero recoil there are no $\Lambda_{{\rm QCD}}/m_c$
corrections to the $m_b, m_c \rightarrow \infty$ predictions of heavy-quark
symmetry for the $B \rightarrow D^*$ and $B \rightarrow D$ matrix
elements of the weak current.$^{[12]}$  Consequently this method for
determining
$|V_{cb}|$ is on a very sound theoretical footing.  Eventually, with
improved data on semileptonic $B$ decay, the error on $|V_{cb}|$
should be substantially reduced.

Experimental information on endpoint of the electron spectrum in
semileptonic $B$-meson decay and $B^0 - \bar B^0$ mixing constrain the
values for $\rho$ and $\eta$.  However, in these cases there are large
theoretical uncertainties that arise from the influence of nonperturbative
strong interaction physics on the relevant hadronic matrix elements.

At the present time the value of $|V_{ub}|$ is determined by comparing
data on the endpoint of the electron spectrum in $B$-meson semileptonic
decay with phenomenological models.  This gives$^{[2]}$ $|V_{ub}/V_{cb}| = 0.10
\pm 0.03$, leading to the constraint $\sqrt{\rho^2 + \eta^2} = 0.5 \pm
0.2$.  Because of the model dependence in extracting $|V_{ub}|$ the error
quoted above should be interpreted as providing a rough measure of the
uncertainty.  In Fig. 4 we have plotted semicircles corresponding to
$\sqrt{\rho^2 + \eta^2} = 0.7$ and $0.3$.  In the future exclusive
decays may also provide valuable information on $|V_{ub}|$.  Heavy quark
symmetry plus isospin symmetry relates the hadronic form factors for $D
\rightarrow \rho \bar e \nu_e$ and $B \rightarrow \rho
e\bar\nu_e$ decay.$^{[8]}$  Since the weak mixing angles are known in the $D$
decay
case the form factors needed for the $B$ decay can be determined from
experimental data on semileptonic $D$ decay.  If light quark $SU(3)$ is
used instead of isospin then the needed decay is $D \rightarrow K^* \bar
e\nu_e$.
There is already experimental information on form factors for this
decay.  Unfortunately there is no theorem that protects the
$m_b, m_c \rightarrow \infty$ heavy quark
symmetry prediction for the relationship between form factors in $B
\rightarrow \rho e\bar\nu_e$ decay and $D \rightarrow \rho \bar e\nu_e$
decay from $\Lambda_{{\rm QCD}}/m_c$  corrections.
Lattice Monte Carlo calculations can also provide valuable information
on the needed hadronic form factors.$^{[13]}$

The measurement $\Delta
M/\Gamma = 0.75$ for $B^0 - \bar B^0$ mixing provides information on the
magnitude of $V_{td}$ (for a given top quark mass).  This constraint
depends on  the hadronic matrix element of a local
four-quark operator, which is usually written as $B_B f_B^2$.  Heavy
quark symmetry$^{[14]}$ and the constituent quark model suggest a
value for $f_B$ around 120 MeV, while recent lattice QCD
results$^{[15]}$ and QCD
sum rule calculations$^{[16]}$ suggest a large value\foot{Calculations
using 2-D QCD in the large $N_c$ limit$^{[17]}$ suggest there are large
$\Lambda_{{\rm QCD}}/m_c$ corrections to the prediction of heavy quark
symmetry for the relationship between $f_D$ and $f_B$.} for $f_B$ around 250
MeV.
Constraining $|V_{td}|$ from $B^0 - \bar B^0$ mixing corresponds to a
constraint on $\sqrt{(1 - \rho)^2 + \eta^2}$, which would appear as a
semicircle in the $\rho - \eta$ plane
centered around $\rho = 1, \eta = 0$.  Because the range of $f_B$'s
mentioned above gives rise to a very large uncertainty in $|V_{td}|$ we
have not plotted the implications of the measured value for $B^0 - \bar
B^0$ mixing in Fig. 4.  Qualitatively the smaller
value $\sqrt{B_B} f_B \simeq$ 120 MeV implies, for top quark masses less
than 200 GeV, that $\rho$ is negative while  the larger value
$\sqrt{B_B} f_B \simeq$ 250 MeV implies that $\rho$ is positive.

Since $Imf(s)$ is small (provided the two photon contribution to the
parity violating amplitude is negligible)  measurement of the polarization
asymmetry $\Delta_{LR}$ determines the value of $\rho$ restricting the
$\alpha$ vertex of the unitary triangle to lie on a vertical line in the
$\rho - \eta$ plane.  Integrating over  the whole available phase space
we find that the interference of the short distance contribution to  the
parity violating amplitude with the parity conserving part implies
that\foot{This differs slightly from the result of Ref. [3] because
in this paper the measured $s$-dependence of $f(s)$ has been used.}
$$      |\Delta_{LR}| =  |2.3 Re \xi| \,\, . \eqno (27)$$
For $m_t =$ 140 GeV and $\rho = - 0.51$ this gives $|\Delta_{LR}| = 3.7
\times 10^{-3}$ while for $m_t =$ 200 GeV and $\rho = - 0.12$ this gives
$|\Delta_{LR}| = 4.7 \times 10^{-3}$.

The magnitude of the asymmetry $\Delta_{LR}$ is larger for $\cos \theta$
positive than for $\cos\theta$ negative as eq. (12) of Section 2
indicates.  Hence, the asymmetry can be increased by a cut on
$\cos \theta$.  If $\cos \theta$ is restricted to lie in the region
$$      - 0.5 < \cos \theta < 1.0 \,\, , \eqno (28)$$
the asymmetry arising from the interference of the short distance
parity violating amplitude with the parity
conserving part is
$$      |\Delta_{LR}| = |4.1 Re \xi| \,\, . \eqno (29)$$
For $m_t =$ 140 GeV, and $\rho = - 0.51$ this gives $|\Delta_{LR}| = 6.5
\times 10^{-3}$ while for $m_t =$ 200 GeV and $\rho = - 0.12$, eq. (29),
implies that $|\Delta_{LR}| = 8.3 \times 10^{-3}$.  This cut increases
the magnitude of the asymmetry by almost a factor of two and  reduces
the number events by only a factor of 0.77.  In Fig. 4 we show the constraint
on $\rho$ extracted from a $\Delta_{LR}$ measurement (with the cut in eq.
(28)) for some  values of the top quark mass and
asymmetry.  The values of the asymmetry and top quark mass are chosen to
be compatible with the measured value for  $B^0 - \bar B^0$ mixing when
$\sqrt{B_B}f_B$ lies between 120 MeV and 250 MeV.  $\xi$ is dominated by the
top quark loop for the values of the  asymmetry shown in Fig. 4.

\noindent {\bf 5.  Two Photon Contribution to the Parity Violating
Amplitude}

In this section we use chiral perturbation theory to examine the two
photon contribution to the parity violating form factors $B$ and $C$.
There are local operators that can contribute to the parity violating
$K^+ \rightarrow \pi^+ \mu^+ \mu^-$ amplitude.  At the leading order of
chiral perturbation theory they are included in the effective
Lagrange density
$$      {\cal L} = {iG_F \alpha s_1 \over \sqrt{2}} \bar\mu \gamma_\mu
\gamma_5 \mu \big[\gamma_1 Tr (O_W Q^2 \Sigma \partial^\mu
\Sigma^{\dagger})$$
$$      + \gamma_2 Tr (O_W \partial^\mu \Sigma Q^2 \Sigma^{\dagger} -
O_W \Sigma Q^2 \partial^\mu \Sigma^{\dagger})$$
$$      + \gamma_3 Tr (O_W \partial^\mu \Sigma Q \Sigma^{\dagger} Q -
O_W \Sigma Q \partial^\mu \Sigma^{\dagger} Q)\big] \,\, . \eqno (30)$$
In eq. (30) $Q$ is the electromagnetic charge matrix
$$      Q = \pmatrix{2/3 & 0 & 0\cr
0 & - 1/3 & 0\cr
0 & 0 & - 1/3\cr} \,\, . \eqno (31)$$
Each term contains two factors of $Q$ because the Lagrange density
in eq. (30) arises from Feynman diagrams with two photons.  When
the photons (and other virtual particles) are off-shell by an amount that is
large compared with the
pseudo-Goldstone boson masses their effects are reproduced by those of
the local operators in eq. (30).  CPS symmetry$^{[18]}$ has been used to reduce
the effective Lagrangian to the form in eq. (30).  Under a CPS
transformation
$$      \Sigma (\vec x, t) \rightarrow S\Sigma^* (- \vec x, t) S \,\, ,
\eqno (32)$$
where $S$ is the matrix that switches strange and down quarks
$$      S = \pmatrix{0 & 0 & 0\cr
0 & 0 & 1\cr
0 & 1 & 0\cr} \,\, . \eqno (33)$$
It is CPS symmetry that forces the two terms in the last two traces of
eq. (30) to occur with a relative minus sign (the linear combination
with a relative plus sign is not invariant under CPS).  Expanding out
the $\Sigma$ matrices in terms of the pseudo-Goldstone boson fields it is
easy to see that the
effective Lagrange density in eq. (30) gives a contribution to $B$
proportional to $\gamma_1 - 8 \gamma_2 - 4\gamma_3$, but gives {\it no}
contribution to $C$.  We shall not be able to predict $B$ using chiral
perturbation theory as $\gamma_1, \gamma_2$ and $\gamma_3$ are not
known.

CPS symmetry forces the contribution to $C$ from local operators
(without factors of $m_q$) to
vanish.  This symmetry is broken by the difference between strange and
down quark masses.  In the pole type graphs of Fig. 5 the quark masses
cannot be neglected and it is  these diagrams that (in chiral
perturbation theory) give the dominant contribution to $C$.  In Fig. 5
the shaded square is an interaction vertex from the weak $\Delta s = 1$
Lagrangian in eq. (18), the shaded circle is a $\eta \gamma \gamma$ or
$\pi^0 \gamma \gamma$ vertex from the Wess-Zumino term$^{[19]}$
$$      {\cal L}_{WZ} = {\alpha\over 4\pi f} \epsilon_{\mu \nu
\lambda\sigma} F^{\mu \nu} F^{\lambda\sigma} (\pi^0/\sqrt{2} +
\eta/\sqrt{6}) + ... \,\, . \eqno (34)$$
The cross denotes a $\eta \mu^+ \mu^-$ or $\pi^0 \mu^+ \mu^-$
vertex that arises from the local terms in the effective Lagrange density
for strong and electromagnetic interactions
$$      {\cal L} = {i\alpha^2\over 4\pi^2} \bar\mu \gamma^\mu \gamma_5
\mu \big[\chi_1 Tr (Q^2 \Sigma^{\dagger} \partial_\mu \Sigma - Q^2
\partial_\mu \Sigma^{\dagger}\Sigma)$$
$$      + \chi_2 Tr (Q \Sigma^{\dagger} Q \partial_\mu \Sigma - Q
\partial_\mu \Sigma^{\dagger} Q \Sigma)\big] \,\, , \eqno (35)$$
that couple a $\pi^0$ or $\eta$ to a $\mu^+ \mu^-$ pair.

In the Feynman diagrams of Fig. 5 the ``infinite part'' of the loop
integrals is cancelled by the terms from eq. (35) yielding the following
prediction for $C$
$$      C = {g_8 \over 12} \left\{{3m_\eta^2 - m_\pi^2 -
2m_{K^{+}}^2\over s - m_\eta^2}\right\} {\cal A}(s) \,\, , \eqno (36)$$
where
$$      Re {\cal A}(s) = {\alpha\over 4\pi^2} \bigg\{w + {1\over 2}
(s/m_\mu^2) - {1\over 4} (s/m_\mu^2)^2$$
$$      + (s/m_\mu^2) ln (s/m_\mu^2) + {1\over 2} (s/m_\mu^2)^2 ln
(s/m_\mu^2)$$
$$      - \int_0^1 dx \left[3 + {2[(s/4m_\mu^2) - 1]
\sqrt{x}\over \sqrt{x + (4 m_\mu^2/s) (1 - x)}}\right]
\lambda_+^2 ln|\lambda_+/2|$$
$$      - \int_0^1 dx \left[ 3 - {2 [(s/4m_\mu^2) - 1]
\sqrt{x}\over \sqrt{x + (4m_\mu^2/s) (1 - x)}} \right]
\lambda_-^2 ln |\lambda_-/2|\bigg\} \,\, , \eqno (37a)$$
and\foot{The imaginary part is related to the unitarity limit for $\eta
\rightarrow \mu^+ \mu^-$.   This was computed in Ref. [20].  The real
part of the $\eta \rightarrow \mu^+ \mu^-$ amplitude was also computed
in Ref. [20] using a phenomenological model for the form factor
associated with the $\eta \rightarrow \gamma \gamma$ vertex.}
$$      Im{\cal A}(s) = {\alpha\over \pi} ~ {1\over \sqrt{1 - (4m_\mu^2/s)}}
ln \left({1 + \sqrt{1 - (4m_\mu^2/s)}\over 2m_\mu/\sqrt{s}}\right) \,\,
. \eqno (37b)$$
The Feynman diagrams in Fig. 5 give no contribution to $B$.  In eq.
(37a) $w$ is a constant independent of $s$ and
$$      \lambda_\pm = \sqrt{x (s/m_\mu^2)} \pm \sqrt{x (s/m_\mu^2)
+ 4 (1 - x)} \,\, . \eqno (38)$$

The constant $w$ gets contributions both from the one loop diagrams and
from the tree diagrams in Fig. 5.  It can be determined from the
relative strength of the decays $\eta \rightarrow \gamma \gamma$ and
$\eta \rightarrow \mu^+ \mu^-$.  At the leading order of chiral
perturbation theory
$$      \Gamma (\eta \rightarrow \gamma \gamma) = {\alpha^2
m_\eta^3\over 96\pi^3 f^2} \,\, , \eqno (39)$$
and
$$      \Gamma (\eta \rightarrow \mu^+ \mu^-) = {|\alpha{\cal A}
(m_\eta^2)|^2\over 48\pi} \left({m_\mu\over f}\right)^2 \sqrt{m_\eta^2 -
4m_\mu^2} \,\, . \eqno (40)$$
The recent measurement$^{[21]}$ of the branching ratio for $\eta \rightarrow
\mu^+ \mu^-, Br (\eta \rightarrow \mu^+ \mu^-) = ( 5 \pm 1) \times
10^{-6}$, is within $1\sigma$ of the unitarity limit which is
$4.3 \times  10^{-6}$ (arising from an on shell two photon intermediate state.)
The measured branching ratio for $\eta \rightarrow \mu^+ \mu^-$ implies
that $|ReA(m_\eta^2)| < 2.5 \times 10^{-3}$ which gives $- 2 < w < 25$.
Using the cut on $cos\theta$, given in eq. (28), we
find that the two photon contribution of the parity violating form
factor $C$, to the asymmetry satisfies, $|\Delta_{LR}| < 1.2 \times
10^{-3}$.  Improving the measurement of the branching ratio for $\eta
\rightarrow \mu^+ \mu^-$ would reduce the uncertainty in $w$ and consequently
improve our knowledge of the two photon contribution to $C$.

If the short distance contribution to the asymmetry $\Delta_{LR}$ (with
the cut on $\cos \theta$ given in eq. (28)) is at the ${1\over 2}\%$ level then
it is likely that the two photon contribution to $C$ can be neglected.
(Of course, if the full range of $\cos \theta$ is used then the
contribution of $C$ to the asymmetry vanishes.)  We have not been able to
predict using chiral perturbation theory, the two photon contribution to
the parity violating form factor $B$.  However, we do not expect its
influence on $\Delta_{LR}$ (with the cut on $\cos \theta$ given in eq.
(28)) to be larger than that of $C$.  (Our naive expectation is that it
gives $|\Delta_{LR}| \sim O (\alpha/\pi) \sim 2 \times 10^{-3}$.)  It
would be interesting to try to estimate the two photon contribution to
$B$ using phenomenological models.  Experimental information on the
decay $K^+ \rightarrow \pi^+ \gamma \gamma$ may also prove useful.

There are $T$-odd asymmetries that involve both the $\mu^+$ and
$\mu^-$ polarizations.  They will be much more difficult to measure than
the parity violating asymmetry we have been discussing.  The $T$-odd
asymmetries also violate parity and are determined by $ImB f^*(s)$ and
$Im C f^*(s)$.  They get a contribution from the interference of the two
photon contribution to the imaginary part of $C$, given in eqs. (36) and
(37),with the real part of the parity conserving amplitude (as well as
from short distance physics).

\noindent {\bf 6.  Concluding Remarks}

We have calculated the two photon contribution to the parity violating
$K^+ \rightarrow \pi^+ \mu^+\mu^-$ decay amplitude arising from the diagrams
in Fig. 5.  They give rise to an invariant matrix element with Lorentz
structure $(p_K - p_\pi)^\mu \bar u \gamma_\mu \gamma_5 v$ and do not
contribute to the other possible form for the parity violating
amplitude, $(p_K + p_\pi)^\mu \bar u \gamma_\mu \gamma_5 v$.  CPS
symmetry of the chiral Lagrangian forces the contact terms (that arise
from Feynman diagrams where the virtual particles have large momentum)
to have the structure $(p_K + p_\pi)^\mu \bar u \gamma_\mu \gamma_5 v$.
Therefore, the diagrams in Fig. 5 give the leading value for the
coefficient of $(p_K - p_\pi)^\mu \bar u \gamma_\mu \gamma_5 v$ in
chiral perturbation theory.  The prediction of chiral perturbation
theory contains an s-independent constant that
is fixed by the measured $\eta \rightarrow \mu^+ \mu^-$ decay rate.
Improving the experimental value for the $\eta \rightarrow \mu^+\mu^-$
branching ratio  would
reduce the uncertainty in this constant and hence improve our prediction
for the coefficient of $(p_K - p_\pi)^\mu \bar u \gamma_\mu \gamma_5
v$.  Unfortunately we cannot compute the coefficient of $(p_K +
p_\pi)^\mu \bar u \gamma_\mu \gamma_5 v$ using chiral perturbation
theory since there are several local contact terms that contribute to it
which we cannot fix experimentally.  These contact terms also contribute
to the $K_L \rightarrow \mu^+ \mu^-$ decay amplitude, but for this
amplitude they enter in a different linear combination than for the $K^+
\rightarrow \pi^+ \mu^+ \mu^-$ matrix element and furthermore the
measured $K_L \rightarrow \mu^+ \mu^-$ branching ratio is not accurate
enough to provide a significant constraint.

If all the available phase space is integrated over then the $(p_K -
p_\pi)^\mu \bar u \gamma_\mu \gamma_5 v$ piece of the parity violating
decay amplitude does not contribute to the parity violating asymmetry
$\Delta_{LR} \equiv (\Gamma_R - \Gamma_L)/(\Gamma_R + \Gamma_L)$.
However, it is advantageous to make the cut, $- 0.5 < \cos\theta < 1$,
since it increases the short distance contribution to the asymmetry by
almost a factor of two and diminishes the number of events by only
a factor of 0.77.  With this cut the measured $\eta \rightarrow \mu^+
\mu^-$ branching ratio implies that the two photon contribution to
$\Delta_{LR}$ from the diagrams in Fig. 5 satisfies $|\Delta_{LR}| < 1.2
\times 10^{-3}$.  This asymmetry is much less than the asymmetry arising
from short distance physics involving virtual top and charm quarks,
provided that $\rho$ is
negative.  For $\rho$ positive, the Feynman diagrams in Fig. 5 may
contribute a non-negligable portion of the asymmetry.  It seems likely to us
that the asymmetry coming from the
two photon contribution to the part of the $K^+ \rightarrow \pi^+ \mu^+
\mu^-$ decay amplitude of the form $(p_K + p_\pi)^\mu \bar u \gamma_\mu
\gamma_5 v$ is not much
larger than that arising from the diagrams in Fig. 5.  Our naive
expectation is that it gives rise to an asymmetry $|\Delta_{LR}| \sim O
(\alpha/\pi) \sim 2 \times 10^{-3}$.  It would be interesting to
estimate this part of the parity violating $K^+ \rightarrow \pi^+ \mu^+
\mu^-$ decay amplitude using phenomenological models.  (Such calculations
may reveal a further physical suppression of this amplitude.)  Experimental
information on the decay $K^+ \rightarrow \pi^+ \gamma \gamma$ could also
be valuable.

The asymmetry $\Delta_{LR}$ can provide information on the unitarity
triangle.  Even an experimental limit at the percent level would provide
interesting information on $\rho$.   This may be within the reach of a
dedicated experiment at existing facilities.$^{[22]}$

Short distance
physics contributes to $T$-odd asymmetries involving both the $\mu^+$
and $\mu^-$ polarizations.  We have found that the imaginary part of the
Feynman diagrams in Fig. 5 (that arises from on shell photons) also
contributes.  This effect should be included in analysis of the
implications of measuring these $T$-odd asymmetries.

\noindent {\bf References}

\item{1.}  M. Kobayashi and T. Maskawa, Prog. Theor. Phys., {\bf 49}
(1973) 65.

\item{2.}  Review of Particle Properties, K. Hikasa, et al, (Particle
Data Group) Phys. Rev., {\bf D45} (1992) 1.

\item{3.}  M. Savage and M.B. Wise, Phys. Lett., {\bf B250} (1990) 151.

\item{4.}  P. Agrawal, et al, Phys. Rev. Lett., {\bf 67} (1991) 537;
P. Agrawal, et al, Phys. Rev., {\bf D45} (1992) 2383.

\item{5.}  C. Alliegro, et al, Phys. Rev. Lett., {\bf 68} (1992) 278.

\item{6.}  G. Ecker, A. Pich and E. deRafael, Nucl. Phys., {\bf B291}
(1987) 692.

\item{7.}  J. Ellis and J. Hagelin, Nucl. Phys., {\bf B217} (1983) 189.

\item{8.}  N. Isgur and M.B. Wise, Phys. Lett., {\bf B232} (1989) 113;
Phys. Lett., {\bf B237} (1990) 527.

\item{9.}  S. Nussinov and W. Wetzel, Phys. Rev., {\bf D36} (1987) 130.

\item{10.}  M.B. Voloshin and M.A. Shifman, Sov. J. Nucl. Phys., {\bf 47}
(1988) 199.

\item{11.}  M. Neubert, Phys. Lett., {\bf B264} (1991) 455.

\item{12.}  M. Luke, Phys. Lett., {\bf B252} (1990) 447.

\item{13.}  For a recent review see:  C. Bernard and A. Soni, BNL-47585
(1992), to appear in Quantum Fields on the Computer, ed. M. Creutz,
World Scientific, 1992.

\item{14.}  M.B. Voloshin and M.A. Shifman, Sov. J. Nucl. Phys., {\bf
45} (1987) 292; H.D. Politzer and M.B. Wise, Phys. Lett., {\bf B206} (1988)
681.

\item{15.}  C. Alexandrou, et al, Phys. Lett., {\bf B256} (1991) 60;
C.R. Allton, el al, Nucl. Phys., {\bf B349} (1991) 105; C. Bernard, et
al, Phys. Rev., {\bf D43} (1991) 2140.

\item{16.}  M. Neubert, Phys. Rev., {\bf D45} (1992) 2451.

\item{17.}  B. Grinstein and P.F. Mende, SSCL-64 (1992) unpublished.

\item{18.}  C. Bernard, et al, Phys. Rev., {\bf D32} (1985) 2343.

\item{19.}  J. Wess and B. Zumino, Phys. Lett., {\bf B37} (1971) 95.

\item{20.}  D.A. Geffen and B.L. Young, Phys. Rev. Lett., {\bf 15}
(1965) 316; C. Quigg and J.D. Jackson, UCRL-18487 (1968) unpublished.

\item{21.}  Reported in:  CERN Courier, May 1992, p. 10.

\item{22.}  Y. Kuno, TR1-PP-92-23, Proceeding of the KEK Workshop On
Rare Kaon Decay Physics, Dec. 10-11 (1991).

\bigskip
\noindent {\bf Figure Captions}

\item{1.}  The unitary triangle.

\item{2.}  Z-penguin and W-box Feynman diagrams that contribute to the
$K^+ \rightarrow \pi^+ \mu^+\mu^-$ decay amplitude.

\item{3.}  Feynman diagrams that give the leading contribution to
$Imf(s)$ in chiral perturbation theory.

\item{4.}  Implications of measurement of the asymmetry $\Delta_{LR}$
for the location of the $\alpha$ vertex of the unitarity triangle.

\item{5.}  Feynman diagrams that give the dominant two photon
contribution to $C$ in chiral perturbation theory.

\bye